\documentclass{article}

\usepackage{spconf}
\usepackage{amsmath}
\usepackage{hyperref}
\usepackage{enumitem}
\usepackage{graphicx}
\usepackage{multirow}
\usepackage{multicol}
\usepackage{xcolor}
\newcommand{\norm}[1]{\left\lVert#1\right\rVert}
\usepackage{soul}
\usepackage{float}
\usepackage{tabularx}
\usepackage{csvsimple}
\usepackage{mathtools}
\usepackage{amssymb}
\usepackage{commath}
\usepackage{subcaption}
\usepackage{cite}
\usepackage{relsize}
\usepackage{romannum}
\usepackage{bm}
\usepackage{stmaryrd}

\def\vec#1{\ensuremath{\bm{{#1}}}}

\title{Prosodic Representation Learning and Contextual Sampling for Neural Text-to-Speech}
\begin{document}
	\name{
		Sri Karlapati, 
		Ammar Abbas, 
		Zack Hodari\sthanks{Work done as an intern at Amazon Research, Cambridge, UK.}, 
		Alexis Moinet, 
		Arnaud Joly,
		Penny Karanasou,
	}
	\nameplus{
		Thomas Drugman
	}

	\address{
		Amazon Research, Cambridge, United Kingdom \\
		*The Centre for Speech Technology Research, University of Edinburgh, United Kingdom
	}

	\maketitle

	\begin{abstract}
		In this paper, we introduce Kathaka, a model trained with a novel two-stage training process for neural speech synthesis with contextually appropriate prosody. In Stage~\Romannum{1}, we learn a prosodic distribution at the sentence level from mel-spectrograms available during training. In Stage~\Romannum{2}, we propose a novel method to sample from this learnt prosodic distribution using the contextual information available in text. To do this, we use BERT on text, and graph-attention networks on parse trees extracted from text. We show a statistically significant relative improvement of $13.2\%$ in naturalness over a strong baseline when compared to recordings. We also conduct an ablation study on variations of our sampling technique, and show a statistically significant improvement over the baseline in each case.
	\end{abstract}

	\vspace{-0.4cm}
	\section{Introduction}
	\vspace{-0.1cm}
		Neural text-to-speech (NTTS) techniques have significantly improved the naturalness of speech produced by TTS systems\cite{wang2017tacotron,shen2018natural,yu2019durian,li2018close,vandenOord2016Wavenet,kalchbrenner2018WaveRNN}. We refer to NTTS systems as a subset of TTS systems that use neural networks to predict mel-spectrograms from phonemes, followed by the use of neural vocoder to generate audio from mel-spectrograms.
			
		In order to improve the prosody\footnote{We use the subtractive definition of prosody from \cite{skerry2018towards}.} of speech obtained from NTTS systems, there has been considerable work in learning prosodic latent representations from ground truth speech\cite{zhang2019learning,skerry2018towards}. These methods use the target mel-spectrograms as input to an encoder which learns latent prosodic representations. These representations are used by the decoder in addition to the input phonemes, to generate mel-spectrograms. The latent representations obtained by encoding a target mel-spectrogram at the sentence level will have information that is not directly available from the phonemes, and by the subtractive definition of prosody, we may claim that these representations capture prosodic information. Several variational\cite{zhang2019learning,karlapati2020copycat,akuzawa2018expressive} and non-variational\cite{skerry2018towards,lee2019robust} methods have been proposed for learning prosodic latent representations. While these methods improve the prosody of synthesised speech, they need an input mel-spectrogram which is not available while running inference on unseen text. This gives rise to the problem of sampling from the learnt prosodic space. Sampling at random from the prior\cite{akuzawa2018expressive} may result in the synthesised speech not having contextually appropriate prosody, as it has no relationship with the text being synthesised. In order to improve the contextual appropriateness of prosody in synthesised speech, there has been work on using textual features like contextual word embeddings and other grammatical information to directly condition NTTS systems\cite{hayashi2019pre,ming2019feature,guo2019exploiting}. These methods require the NTTS model to learn an implicit correlation between the given textual features and the prosody of the sentence. One work also poses this sampling problem as a selection problem and uses both syntactic distance and BERT embeddings to select a latent prosodic representation from the ones seen at training time\cite{tyagi2019dynamic}.

		Bringing both the aforementioned ideas of using ground truth speech to learn prosodic latent representations and using textual information, we build Kathaka, a model trained using a two-stage training process to generate speech with contextually-appropriate prosody. In Stage~\Romannum{1}, we learn the distribution of sentence-level prosodic representations from ground truth speech using a VAE\cite{zhang2019learning}. In Stage~\Romannum{2}, we learn to sample from the learnt distribution using text. In this work, we introduce the BERT+Graph sampler, a novel sampling mechanism which uses both contextual word-piece embeddings from BERT\cite{devlin2019bert} and the syntactic structure of constituency parse trees through graph attention networks\cite{velivckovic2018graph}. We then compare Kathaka against a strong baseline and show that it obtains a relative improvement of $13.2\%$ in naturalness.
		\vspace{-0.5cm}
	
	\section{NTTS Baseline with Duration Modelling}
		\begin{figure*}
			\centering
			\begin{subfigure}{0.61\linewidth}
				\includegraphics[width=\linewidth]{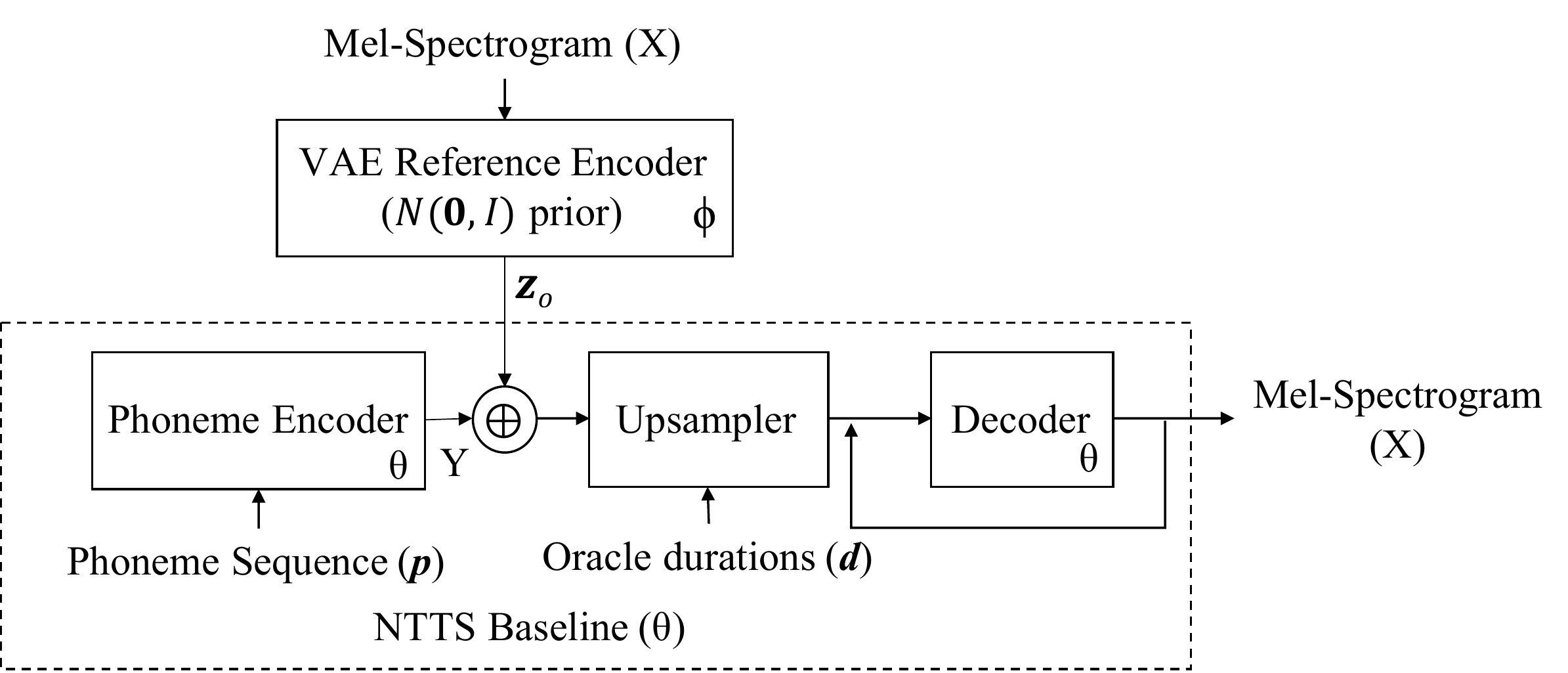}
				\caption{NTTS Model with VAE Reference Encoder}
				\label{fig:vae_model}
			\end{subfigure}
			\hspace*{\fill} 
			\begin{subfigure}{0.34\linewidth}
				\includegraphics[width=\linewidth]{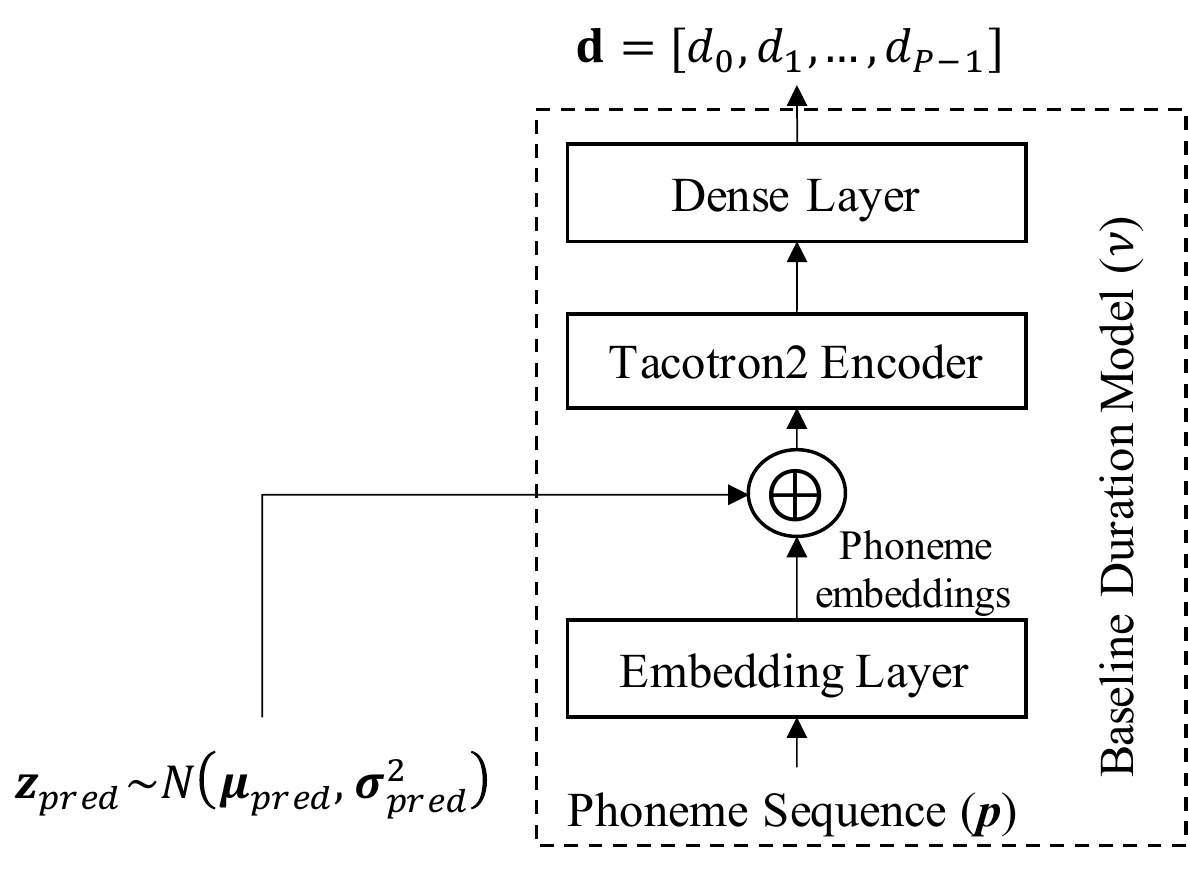}
				\caption{Prosody-dependent Duration Model}
				\label{fig:prosody_dependent_duration_model}
			\end{subfigure}
			\vspace{-0.2cm}
			\caption{Fig.~(a) shows the architecture of the NTTS Baseline system ($\theta$) within the bounding box and the added VAE reference encoder ($\phi$) used in \textbf{Stage~\Romannum{1}} of Kathaka's training procedure. Fig.~(b) shows the architecture of the baseline duration model within the bounding box and the  prosody-dependent duration model proposed in Section~\ref{ssec:prosody_dependent_duration_modelling}.}
			\vspace{-0.4cm}
		\end{figure*}
		\label{sec:baseline}
		We use a modified version of DurIAN\cite{yu2019durian} as our baseline. The baseline learns to predict $T$ mel-spectrogram frames $X=[\vec{x}_0, \vec{x}_1, \dots, \vec{x}_{T-1}]$, given $P$ phonemes $\vec{p}=[p_0, p_1, \dots, p_{P-1}]$. The baseline implicitly models the prosody of synthesised speech without any additional context or reference prosody. We have a phoneme encoder, which takes phonemes as input and returns phoneme encodings $Y=[\vec{y}_0, \vec{y}_1, \dots, \vec{y}_{P-1}]$. At training, we use forced alignment to extract durations, $\vec{d}=[d_0, d_1, \dots, d_{P-1}]$, where $\sum_{d_i \in \vec{d}} d_i=T$. The phoneme encodings are upsampled by replication as per the durations $\vec{d}$ to obtain upsampled phoneme encodings $Y^\uparrow = [\vec{y}^\uparrow_{0}, \vec{y}^\uparrow_{1}, \dots, {\vec{y}^\uparrow_{T-1}}]$. The decoder auto-regressively learns to predict the mel-spectrogram $X$ from $Y^\uparrow$. Unlike DurIAN, we do not use a post-net as this resulted in instabilities when training with a reduction factor of $1$. Our baseline is shown within the bounding box in Fig.~\ref{fig:vae_model}, and the set $\{\theta\}$ represents its parameters.
		
		We train a duration model $\vec{\hat{d}} = f_{\nu}(\vec{p})$, to predict the duration $\hat{d}_{p}$ in frames for each phoneme $p$, as shown by the blocks within the bounding box in Fig.~\ref{fig:prosody_dependent_duration_model}. We note that the distribution of phoneme durations contains multiple modes $M$. We normalize durations for a group of tokens $G_m$, for each mode $m \in M$ separately using mean $\mu_m$ and standard deviation $\sigma_m$. We use L2 loss to train the model as shown below, where $\llbracket \cdot \rrbracket$ represent Iverson brackets.
		\begin{equation}
			\begin{split}
				\mu_m = & \frac{1}{\mid G_m\mid} \sum_{p \in G_m} d_p,\ 
				\sigma_m^2 = \frac{1}{\mid G_m\mid} \sum_{p \in G_m} (d_p - \mu_m)^2 \\
				L_{\nu} &= \frac{1}{{P}}  \sum_{p \in \vec{p}} \norm{\hat{d}_{p} - \prod_{m \in M} \Bigg(\frac{d_p-\mu_m}{\sigma_m}\Bigg)^{\llbracket p \in G_m \rrbracket} }_2
			\end{split}
			\label{eq:dur_pred}
			\vspace{-0.2cm}
		\end{equation}
		\vspace{-1cm}
	\section{Kathaka}
		\vspace{-0.1cm}
		\label{sec:method}
		Here, we introduce the two-stage approach to train Kathaka. In Section~\ref{ssec:vae_reference_encoder}, we describe Stage~\Romannum{1} where the model learns a distribution over the prosodic latent space. In Section~\ref{ssec:sampling_using_text} we introduce our sampling mechanisms and finally discuss prosody dependent duration modelling in Section~\ref{ssec:prosody_dependent_duration_modelling}.
		\vspace{-0.3cm}
		\subsection{Addition of a Variational Reference Encoder}
			\label{ssec:vae_reference_encoder}
			We add a variational reference encoder $q_{\phi}$, with parameters $\{\phi\}$, to the NTTS architecture $\{\theta\}$ as shown in Fig.~\ref{fig:vae_model}, to learn prosodic representations from speech\cite{zhang2019learning}. This encoder $q_{\phi}(\vec{z}_{o}\mid X)$ takes a mel-spectrogram $X$ as input and predicts the parameters of a Gaussian distribution $\mathcal{N}(\vec{\mu}_{o}, \vec{\sigma}^{2}_{o})$, from which we sample a prosodic latent representation $\vec{z}_{o}$. We assume a prior distribution $p_{\theta}(\vec{z}_{o}) = \mathcal{N}(\vec{0}, I)$ and train the model to maximize the evidence lower bound (ELBO) defined in Eq.~\ref{eq:elbo}, where $\alpha$ is used as the annealing factor to avoid posterior collapse\cite{sonderby2016train}.
			\begin{equation}
				\begin{split}
					L_{\phi, \theta} =\ & \mathbb{E}_{\vec{z}_{o} \sim q_{\phi}}[log(p_{\theta}(X \mid \vec{z}_{o}, Y_{\uparrow}))] \\ 
					&\ - \alpha \ D_{KL}(q_{\phi}(\vec{z}_{o}\mid X) \mid\mid p_{\theta}(\vec{z}_{o}))
                    \label{eq:elbo}
                \end{split}
			\end{equation}	
		\subsection{Sampling Using Text}
			\label{ssec:sampling_using_text}
			During inference, we need to generate the mel-spectrogram $X$, and therefore will not have access to it. We note that prosody is driven by the contextual information available in a sentence \cite{wagner2010experimental}. Therefore, we propose the usage of text or features derived from it to learn to predict a sentence-level prosodic latent representation $\vec{z}_{pred}$. This is used in place of a mel-spectrogram based sentence-level prosodic latent representation $\vec{z}_{o}$. We define samplers $s_{\chi}(\vec{z}_{pred}\mid W)$, which take text or textual features $W$ as input, and learn to predict a sentence-level prosodic latent representation $\vec{z}_{pred}$. Since the reference encoder $q_{\phi}$ predicts the parameters of a Gaussian distribution with a diagonal covariance matrix $\mathcal{N}(\vec{\mu}_{o}, \vec{\sigma}^{2}_{o})$, we treat this as a distribution matching problem, and train samplers to predict parameters of a Gaussian distribution $\mathcal{N}(\vec{\mu}_{pred}, \vec{\sigma}^{2}_{pred})$. We match these two distributions by minimising the KL Divergence:
			\begin{equation}
				\label{eq:elbo_expanded}
				\begin{split}
					D_{KL}(s_{\chi}(&\vec{z}_{pred}\mid W) \mid\mid q_{\phi}(\vec{z}_{o}\mid X)) = \\
						\frac{1}{2} &\Bigg(\sum_{i=0}^{D-1} \bigg(log({\vec{\sigma}^2_{o_{i}}}) - log({\vec{\sigma}^2_{{pred}_{i}}}) \\ 
						&+ \frac{\vec{\sigma}^2_{pred_{i}}}{\vec{\sigma}^2_{o_{i}}} + \frac{(\vec{\mu}_{o_{i}} - \vec{\mu}_{pred_{i}})^2}{\vec{\sigma}_{o_{i}}^2}\bigg) - D\Bigg),
                \end{split}
			\end{equation}
			where $D$ is the number of dimensions of the latent representation. During inference, we replace $\vec{z}_{o}$ from the reference encoder $q_{\phi}$, by $\vec{z}_{pred}$ obtained from the sampler $s_{\chi}$. Now we dive into various sampler architectures.
			\vspace{-0.3cm}
			\subsubsection{BERT Sampler}
				\vspace{-0.1cm}
				\label{sssec:bert_sampler}
				BERT is a masked language model known to capture contextual information in a given sentence \cite{devlin2019bert}. This contextual information is captured in the word-piece embeddings provided by BERT for a given text. Since there has been a lot of work showing correlations between the contextual information captured by BERT and the semantics of the sentence\cite{rogers2020primer}, we use BERT as a semantic sampler. As shown in Fig.~\ref{fig:bert_sampler}, we use a BERT model pre-trained on long-form text to get contextual word-piece embeddings from the text. These word-piece embeddings are passed through a bidirectional LSTM. We concatenate the first and last hidden states to get a single sentence-level representation which is then projected to obtain $\vec{\mu}_{pred}$ and $\vec{\sigma}^2_{pred}$. The sampler's parameters are denoted as $\{\gamma\} \subseteq \{\chi\}$, and are trained using the loss in Eq.~\ref{eq:elbo_expanded}, while also fine-tuning BERT at a low learning rate.
				\vspace{-0.1cm}
			\subsubsection{Graph Sampler}
				\vspace{-0.1cm}
				\label{sssec:gnn_sampler}
				Constituency parse trees have long been used to represent the grammatical structure of a piece of text\cite{guo2019exploiting}, and their correlation to prosody is well known\cite{kohn2018empirical}. As shown in Fig.~1 from \cite{tyagi2019dynamic}, these parse trees $Tr$ have the words in a sentence as their leaves, and each leaf has only one parent node which represents the part-of-speech of that word. Upon removing the word nodes, the tree $Tr^{'}$ with it's internal nodes can be considered as a representation of the syntax of the sentence. Since all trees $Tr^{'}$ can be represented as undirected acyclic graphs $G$, we propose to use the tree as input to a Graph Neural Network. Graph Neural Networks have been used to learn representations at the node level, while exploiting the structure of the input \cite{li2015gated,zhou2018graph,velivckovic2018graph}. We train a Message-Passing based Graph Attention Network (MPGAT)\cite{velivckovic2018graph} with one attention head, which generates a node level representation based on the structure of the tree. In one pass of messages between connected nodes, a representation is learnt at every node depending on the connected nodes and itself. We pass $N$ such messages, where $N$ is the 75th percentile of the distribution of graph diameters, so that every node also has information pertaining to nodes that are not its immediate neighbours. As shown in Fig.~\ref{fig:graph_sampler}, we extract $G$ from the text, and pass it through MPGAT. Upon obtaining a representation at each node, we extract the representations at the leaves in a depth-first order in order to obtain the representations at the parts-of-speech nodes in relation to the structure of the tree. We pass them through a bidirectional LSTM, concatenate the first and last hidden states to get a sentence-level representation, and project it to obtain $\vec{\mu}_{pred}$ and $\vec{\sigma}^2_{pred}$. The sampler's parameters are denoted as $\{\omega\} \subseteq \{\chi\}$, and are trained using the loss in Eq.~\ref{eq:elbo_expanded}.
				\vspace{-0.3cm}
			\subsubsection{BERT+Graph Sampler}
\begin{figure}
					\centering
					\begin{subfigure}{\linewidth}
						\includegraphics[width=\linewidth]{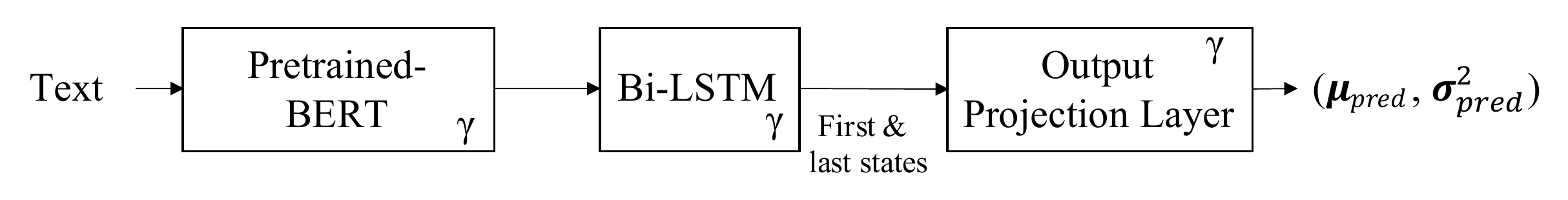}
						\caption{BERT Sampler($\{\gamma\}$)}
						\label{fig:bert_sampler}
					\end{subfigure}
					\begin{subfigure}{\linewidth}
						\includegraphics[width=\linewidth]{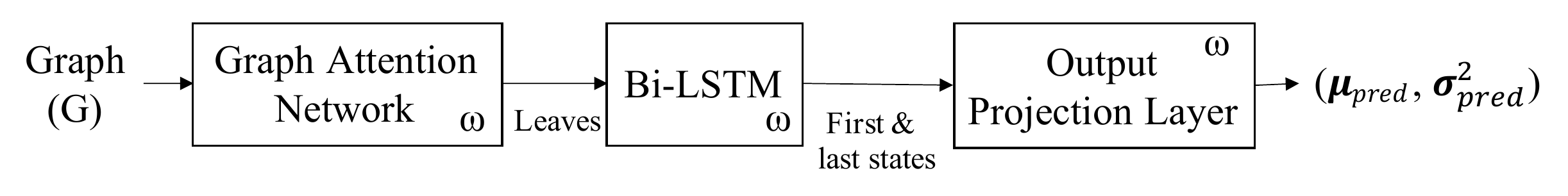}
						\caption{Graph Sampler($\{\omega\}$)}
						\label{fig:graph_sampler}
					\end{subfigure}
					\begin{subfigure}{\linewidth}
						\includegraphics[width=\linewidth]{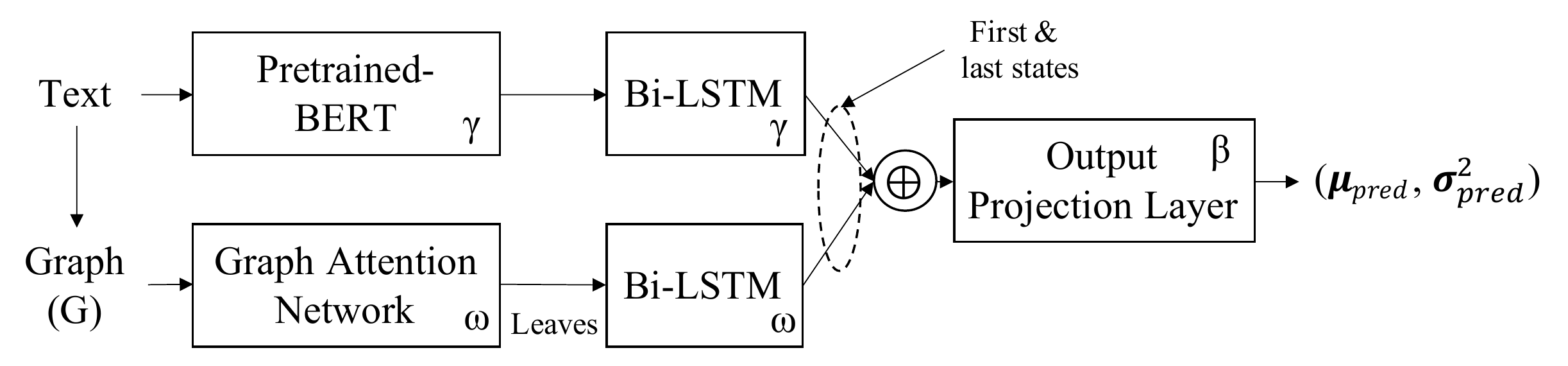}
						\caption{BERT+Graph Sampler($\{\gamma\} \cup \{\omega\} \cup \{\beta\}\}$)}
						\label{fig:bert_graph_sampler}
					\end{subfigure}
					\vspace{-0.3cm}
					\caption{Fig.~(a) and (b) show the architectures of the BERT and Graph Samplers. Fig.~(c) shows the architecture of the BERT+Graph Sampler used in \textbf{Stage~\Romannum{2}} of training Kathaka.}
					\vspace{-0.3cm}
				\end{figure}

				\vspace{-0.12cm}
				\label{sssec:bert_gnn_sampler}
				In Fig.~\ref{fig:bert_graph_sampler}, we combine the BERT and Graph samplers by concatenating the sentence-level representations obtained from each of the methods, and projecting them to obtain $\vec{\mu}_{pred}$ and $\vec{\sigma}^2_{pred}$. We denote the set of all parameters in this sampler as ${\{\gamma\} \cup \{\omega\} \cup \{\beta\}} \subseteq \{\chi\}$, where $\{\beta\}$ is the set of projection parameters, and we train using Eq.~\ref{eq:elbo_expanded}.
				\vspace{-0.3cm}
		\subsection{Prosody-dependent duration modelling}
			\label{ssec:prosody_dependent_duration_modelling}
			\vspace{-0.12cm}
			We condition our duration model on the phoneme embeddings and the sampled prosodic latent representation, $\vec{z}_{pred}\ \sim \mathcal{N}(\vec{\mu}_{pred}, \vec{\sigma}^2_{pred})$. Thus, as shown in Fig.~\ref{fig:prosody_dependent_duration_model}, we are modelling $\vec{d} = f_{\nu}(\vec{p}, \vec{z}_{pred})$.
			\begin{figure*}
				\centering
				\begin{subfigure}{0.49\linewidth}
					\includegraphics[width=\linewidth]{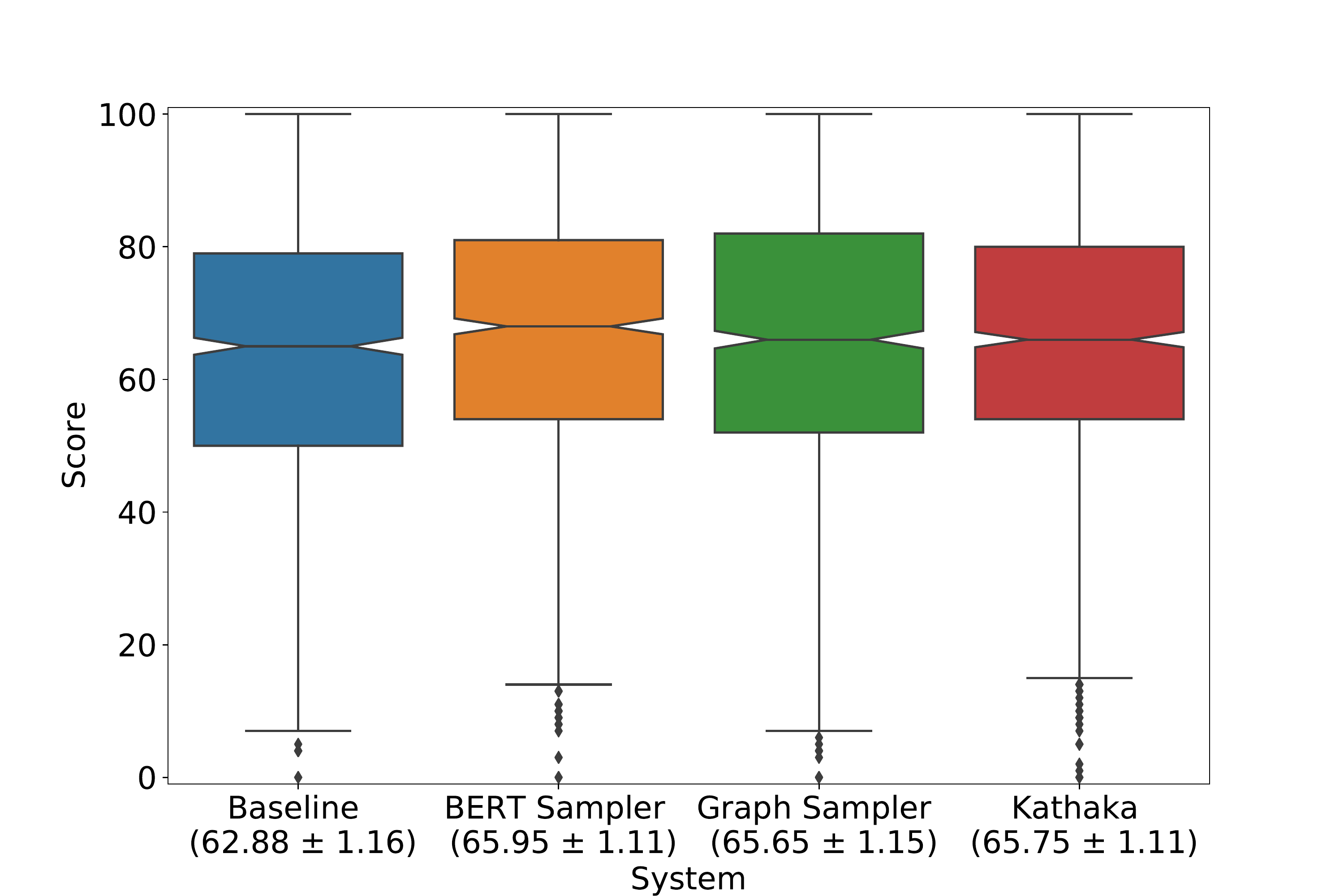}
					\caption{MUSHRA Scores from Ablation Study}
					\label{fig:ablation_mushra}
				\end{subfigure}
				\begin{subfigure}{0.49\linewidth}
					\includegraphics[width=\linewidth]{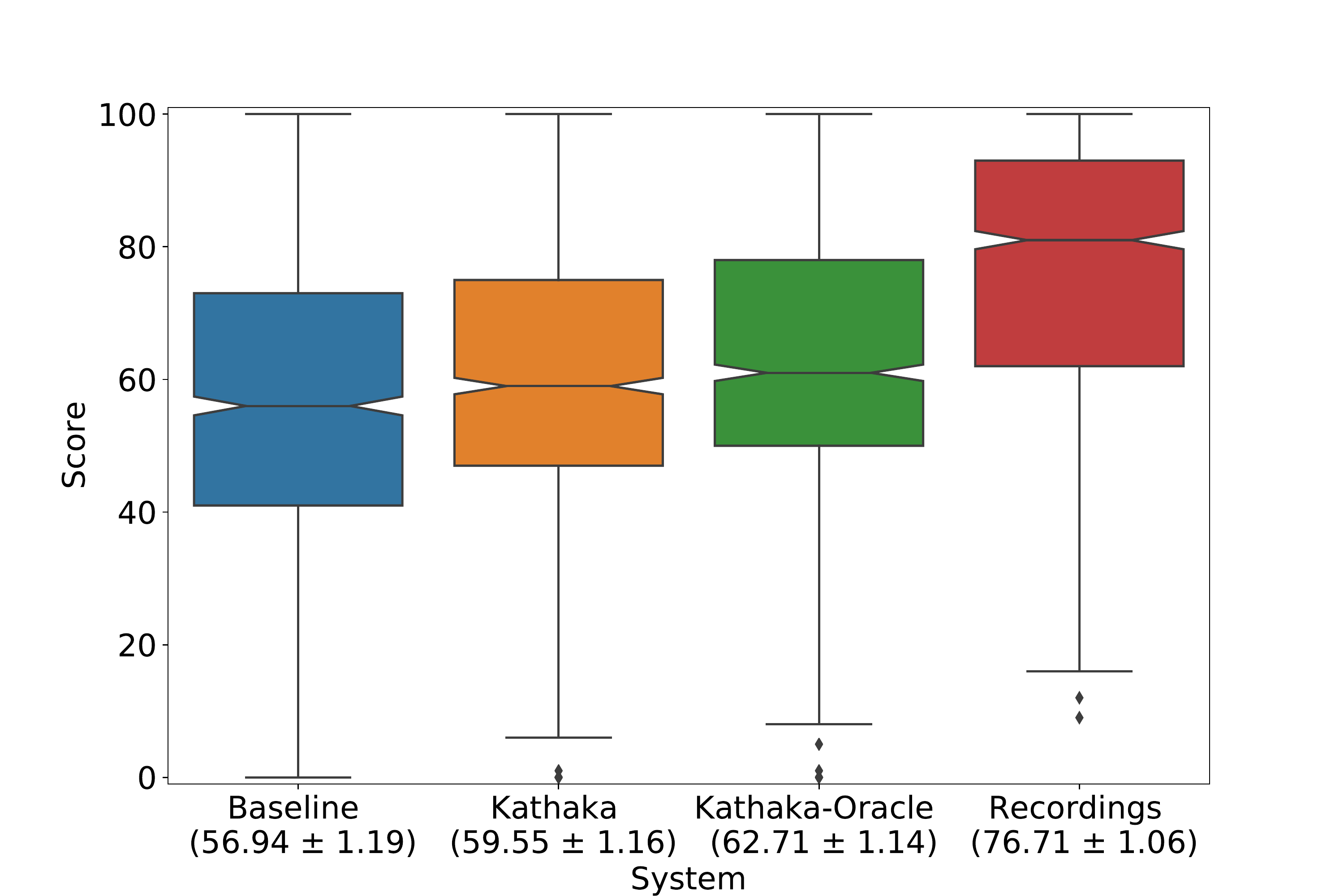}
					\caption{MUSHRA Scores from Gap Reduction Study}
					\label{fig:baseline_comp_mushra}
				\end{subfigure}
				\vspace{-0.12cm}
				\caption{In both figures, we show the box plots of the scores obtained by each of the systems in the test. We show the mean score with $95\%$ confidence interval range next to the system names. In Fig.~(a), all the samplers are statistically significantly better than the baseline. In Fig.~(b), Kathaka obtains a statistically significant relative improvement of $13.2\%$ over the baseline.}
			\end{figure*}
			\vspace{-0.3cm}
		\subsection{Training \& Inference}
			\vspace{-0.12cm}
			\label{ssec:training_setup}
			Our model is trained in 3 steps: 1)~we train the model with a variational reference encoder and durations obtained from forced alignment as in Section~\ref{ssec:vae_reference_encoder}; 2)~we train a sampler on text as in Section~\ref{ssec:sampling_using_text}; 3)~we train a prosody-dependent duration model as in Section~\ref{ssec:prosody_dependent_duration_modelling} using the prosodic latent space $\mathcal{N}(\vec{\mu}_{pred}, \vec{\sigma}^2_{pred})$ learnt in Step 2. During inference, we perform 3 steps: a)~we use the sampler to predict the latent vector $\vec{z}_{pred}$; b)~we concatenate $\vec{z}_{pred}$ obtained in Step~a, with the phoneme embeddings $Y$, to predict durations; c)~we use upsampled phoneme encodings $Y^{\uparrow}$, $\vec{z}_{pred}$, and the predicted durations, to get mel-spectrograms using the model trained in Step~1 of training. Speech is synthesised from mel-spectrograms by using a WaveNet vocoder~\cite{vandenOord2016Wavenet}.
			\vspace{-0.15cm}
	\section{Experiments}
		\label{sec:experiments}
		\vspace{-0.3cm}
		\subsection{Data}
			\vspace{-0.15cm}
			\label{ssec:data}
			We used 38.5 hours of an internal long-form US English dataset recorded by a female narrator. 33 hours of this dataset was used as the training set and the remainder as the test set.
			\vspace{-0.15cm}

		\subsection{Evaluation}
			\label{ssec:evaluation}
			We conducted 2 separate MUSHRA\cite{itu20031534} tests for evaluating the naturalness of our system and the impact of different sampling techniques. Both MUSHRA tests were taken by 25 native US English listeners. Each test consisted of 50 samples which were 15 seconds in length on average. The listeners rated each system on a scale from 0 to 100 in terms of naturalness. We evaluated statistical significance using pairwise two-sided Wilcoxon signed-rank tests.
			
			\subsubsection{Ablation Study}
				\label{sssec:ablation_study}
				We compared each of the samplers mentioned in Section~\ref{ssec:sampling_using_text}, with the baseline in a 4-system MUSHRA. Each of the samplers showed a statistically significant (all systems had a $\text{p-value}<10^{-3}$, when compared to the baseline) improvement over the baseline, as seen from Fig.~\ref{fig:ablation_mushra}. There is no statistical significance in the differences in the scores between the samplers themselves. This shows that both the BERT and Graph samplers may be equally capable at the task of sampling from the learnt latent prosodic space, therefore, even upon combining them, there is no statistically significant difference in the MUSHRA scores of the samplers. We selected the BERT+Graph sampler as the sampler for the Kathaka model, and used it to measure the reduction in gap as it had the lowest minimum score among the three samplers.
			
			\subsubsection{Gap Reduction Study}
				\label{sssec:gap_reduction}
				In this MUSHRA test, we evaluated 4 systems, namely: 1)~the baseline, 2)~Kathaka, 3)~Kathaka-Oracle, the VAE based NTTS model with oracle prosodic embeddings obtained from recordings, and 4)~the original recordings from the narrator. Kathaka obtained a relative improvement over the baseline by a statistically significant $13.2\%$ ($\text{p-value}<10^{-3}$), as shown in Fig.~\ref{fig:baseline_comp_mushra}. The model using oracle prosodic representations, Kathaka-Oracle, reduces the gap to recordings by $29.19\%$ ($\text{p-value}<10^{-4}$). We hypothesise that the gap between Kathaka and Kathaka-Oracle is due to the sampler being trained at the sentence level. A sentence may not contain all the context required to determine the appropriate prosody with which a piece of text should be rendered, especially in long-form speech synthesis. While using a sampler trained at the sentence level shows a significant relative improvement, we speculate that exploiting the context available beyond the sentence will help to further reduce this gap and is left as future work.
				\vspace{-0.2cm}
	\section{Conclusion}
		\vspace{-0.2cm}
		\label{sec:conclusion}
		We presented Kathaka, an NTTS model trained using a novel two-stage training approach for generating speech with contextually appropriate prosody. In the first stage of training, we learnt a distribution of sentence-level prosodic representations. We then introduced a novel sampling mechanism of using trained samplers to sample from the learnt sentence-level prosodic distribution. We introduced two samplers, 1)~the BERT sampler which uses contextual word-piece embeddings from BERT and 2)~the Graph sampler where we interpret constituency parse trees as graphs and use a Message Passing based Graph Attention Network on them. We then combine both these samplers as the BERT+Graph sampler, which is used in Kathaka. We also modify the baseline duration model to incorporate the latent prosodic information. We conducted an ablation study of the samplers and showed a statistically significant improvement over the baseline in each case. Finally, we compared Kathaka against a baseline, and showed a statistically significant relative improvement of $13.2\%$.
	\newpage
	\bibliographystyle{IEEEbib}
	\bibliography{references}

\end{document}